# In-Pocket 3D Graphs Enhance Ligand-Target Compatibility in Generative Small-Molecule Creation

Seung-gu Kang*, Jeffrey K. Weber†, Joseph A. Morrone†, Leili Zhang,

Tien Huynh, and Wendy D. Cornell

Computational Biology Center, IBM Thomas J. Watson Research

1101 Kitchawan Road, Yorktown Heights, NY 10594, USA

**ABSTRACT:** Proteins in complex with small molecule ligands represent the core of structure-based drug discovery. However, three-dimensional representations are absent from most deep-learning-based generative models. We here present a graph-based generative modeling technology that encodes explicit 3D protein-ligand contacts within a relational graph architecture. The models combine a conditional variational autoencoder that allows for activity-specific molecule generation with putative contact generation that provides predictions of molecular interactions within the target binding pocket. We show that molecules generated with our 3D procedure are more compatible with the binding pocket of the dopamine D2 receptor than those produced by a comparable ligand-based 2D generative method, as measured by docking scores, expected stereochemistry, and recoverability in commercial chemical databases. Predicted protein-ligand contacts were found among highest-ranked docking poses with a high recovery rate. This work shows how the structural context of a protein target can be used to enhance molecule generation.

Recent advances in deep-learning technology promise a new era of data-driven drug discovery, carrying high hopes for the acceleration of lead finding and lead optimization pipelines.[1–3] Small molecule generative models have been designed based on a range of deep-learning architectures,[4,5] often with the intent to produce molecules with desired physicochemical properties in a way that both enhances the chemist's creativity and extends the scale of conventional data- or physics-based approaches.[6,7] Many such models employ SMILES[8] as a principle molecular representation, as molecular strings can bootstrap from the success of existing natural language processing (NLP) technologies.[9] However, despite their suitability for high throughput computational processing, 1-dimensional (1D) strings provide far from physically intuitive representations of small molecules:[10–13] in reality, small molecules are built from intricate networks of interatomic connections that exist in three-dimensional conformational space. While sophisticated text-based deep learning architectures like transformers can sometimes learn spatial relationships from simple strings,[14] architectures encoding representations of the bonded and 3-dimensional (3D) nature of molecules offer a more natural approach to capturing chemistry and physics in deep learning. Accordingly, several recent studies have employed 2-dimensional (2D) graph representations of small molecules in generative models,[15,16] demonstrating success in the functional modulation of generated output.[17,18] Ligand binding is best understood, however, in the context of 3D receptor structure and intermolecular contacts within protein-ligand binding sites.[19–22] A number of recent generative modeling efforts have incorporated direct information from target binding sites, either using genomic[23,24] or protein sequence information[25,26] or more concrete structural data represented with 3D voxels, spatial context, or molecular fragments.[20,27–30]

Here, we present generative models that explicitly encode 3D protein-ligand complexes as molecular graphs, extending a relational graph structure from previous ligand-only graph generative models (Fig 1A).[17] Beyond being physically intuitive and translationally and rotationally invariant, our graph representations are easily extensible to most types of intra- and intermolecular interactions, allowing for straightforward inclusion of interaction partners like proteins and the subsequent prediction of binding mode. Specifically, ligand bonds and their cognate receptor contacts are captured within a global relational graph; such bonds and contacts are treated independently according to specific chemical bond and interaction types corresponding to different edge tensors. Taking these molecular and contact graphs as input, our 3D generative model is built on a conditional variational autoencoder architecture (cVAE) composed of a graph convolutional encoder and RNN-assisted graph decoder. Models are trained via optimization of three objective functions: (1) the reconstruction losses for a ligand's molecular graph and protein contact graph, (2) the regularizing Kullback–Leibler divergence (KLD) between the Gaussian prior and the encoding posterior, and (3) a Jensen-Shannon divergence (JSD) capturing the fidelity of generated protein-ligand interactions (see SI for details).

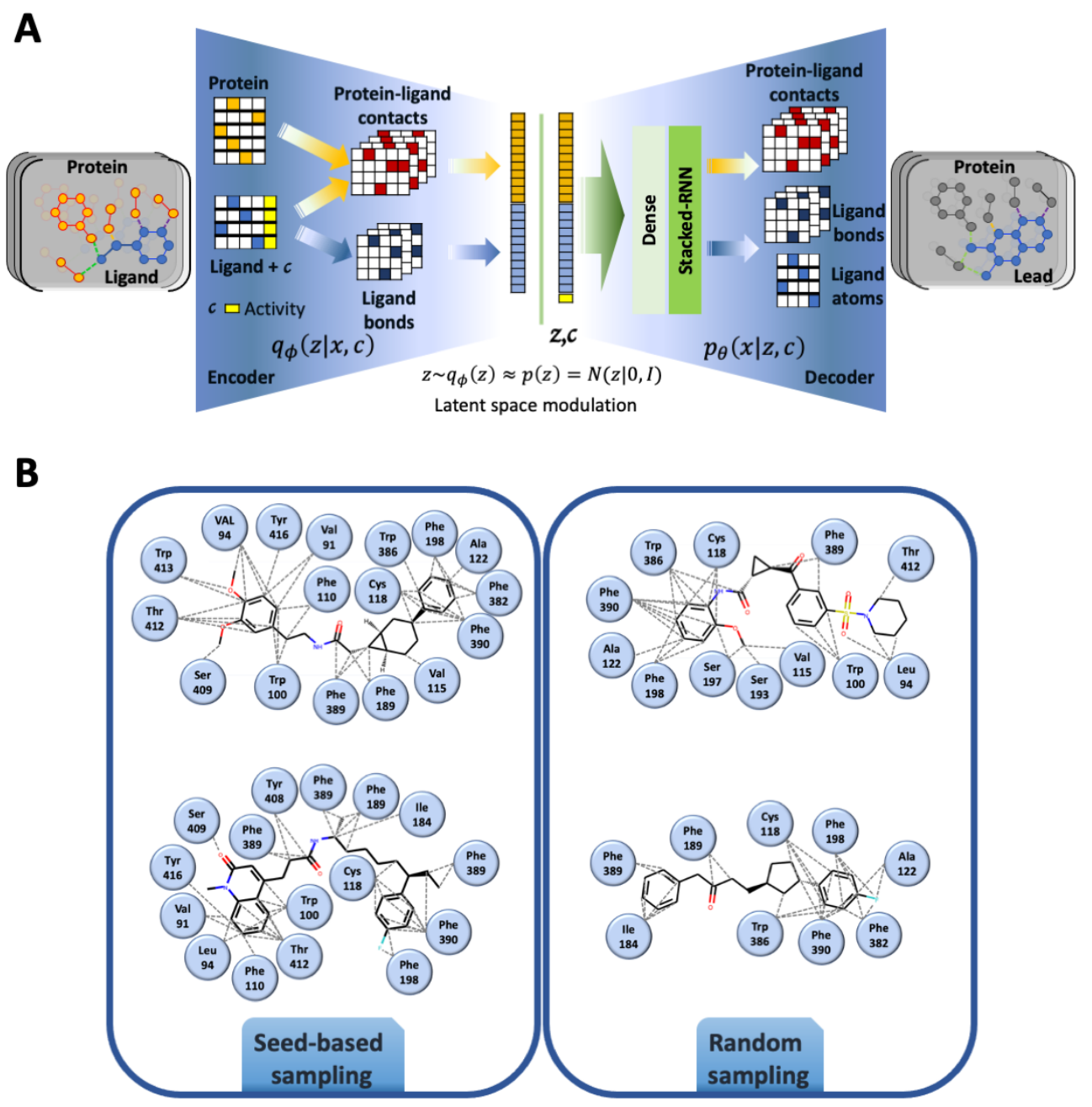

**Figure 1**. Architecture of 3D graph-based generative models, including the 3D protein-ligand interaction network (A) and examples of generated molecules with predicted protein-ligand interactions (dashed lines) generated from seed-based and random sampling (B).

To identify advantages of our 3D approach, we trained both 3D and 2D generative models on a common target (the dopamine D2 receptor – DD2R), common training and test sets, and, to the extent possible, on a common graph cVAE architecture. Input structures for the 3D generative model (gen3D) were prepared by docking a refined PubChem chemical library[31] (with ~1,500 active and ~150,000 inactive molecules) into the crystal structure of DD2R (PDBID: 6CM4). Novel molecules and their putative binding modes were then generated with two different protocols, for both inactive and active conditions: (1) random generation, via direct random sampling of the latent space, and (2) seed-based generation, via random sampling in the latent space neighborhoods of encoded seeds. Training protocols for the 2D generative model (gen2D) are given in Ref.[17].

Examples of generated molecules and their predicted residue-level protein contacts are shown in Fig 1B. A principal question concerns whether explicit encoding of 3D protein-ligand contacts impacts molecular generation relative to gen2D, especially with respect to producing better binding modes. To address this issue, we performed massive docking simulations for molecules generated from both gen3D and gen2D protocols, enumerating all possible enantiomers when stereocenters were present. Each set contained between ~5,000 to ~40,000 molecules depending on specific activity conditioning and generation procedures.

As Fig. 2A illustrates, the seed-based gen3D protocol produced top-pose docking scores nearly a whole energy unit lower, on average (~0.8 kcal/mol) than gen2D, with the same trend holding for the second and third poses (Fig 2A). The activity-conditioning procedure also yielded results more consistent with docking within the gen3D protocol versus gen2D: the active and inactive classes were differentiated by scores at -1.0 vs. -0.4 kcal/mol, respectively. Considering the size of our docking evaluation dataset, our findings strongly support the hypothesis that training with 3D protein contacts improves complementarity between generated molecules and the intended target pocket, as measured by docking scores.

By contrast, the random sampling procedure resulted in statistically comparable docking scores between the gen3D and gen2D protocols, both absolutely and between active and inactive conditioned molecules. This lack of 3D advantage in the random sampling case may result from the increased complexity of the 3D latent space, which requires higher dimensionality to facilitate protein-ligand contact encoding. Although 3D latent-space sampling presently seems to be less effective without the guidance of molecular seeds, seed-guided improvements derived from 3D structure are striking, particularly as sampling around molecular seeds is a common strategy in the generative modeling field and supports lead optimization.

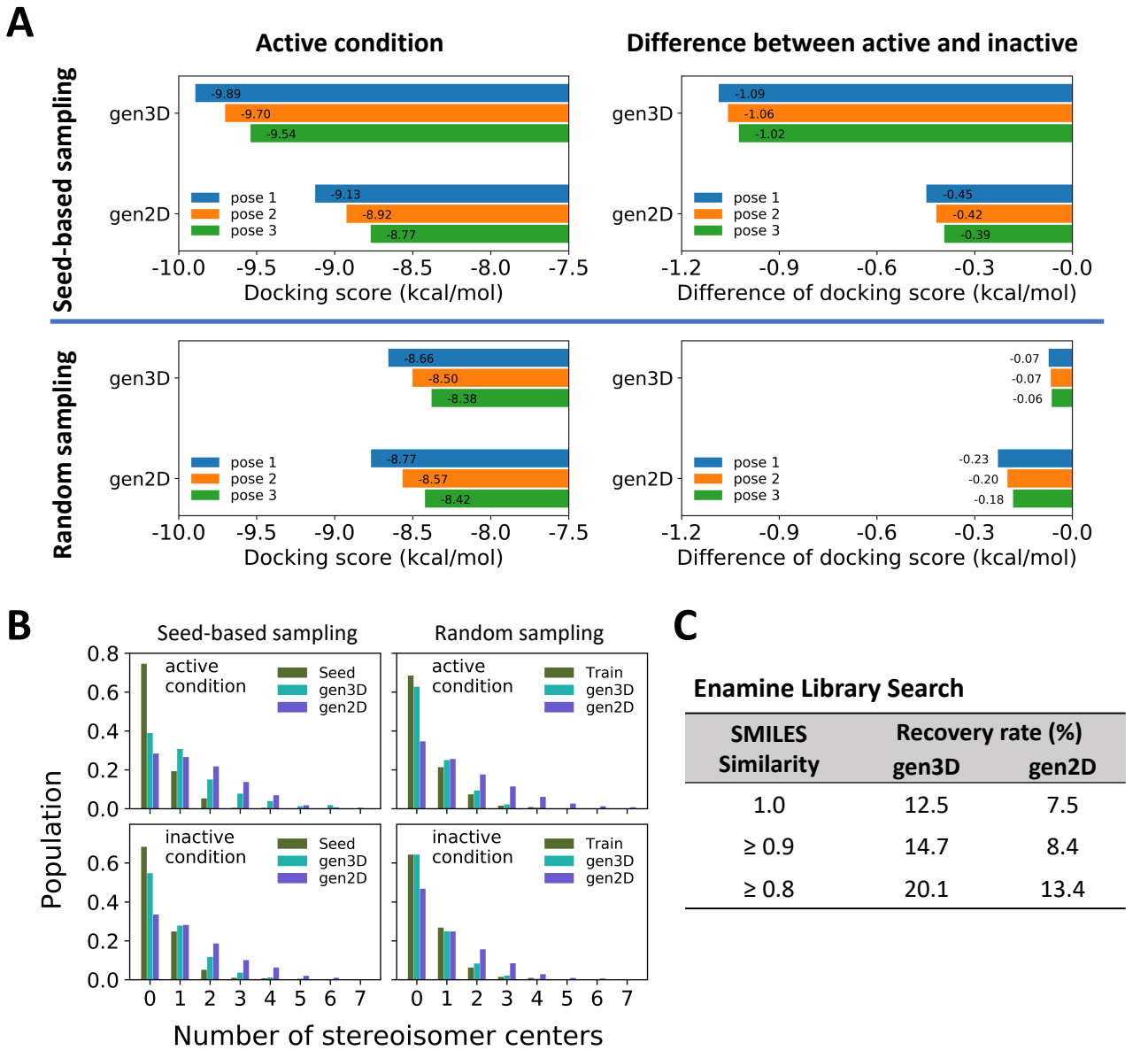

C

**Enamine Library Search**

| SMILES Similarity | Recovery rate (%) gen3D | gen2D |
|---|---|---|
| 1.0 | 12.5 | 7.5 |
| ≥ 0.9 | 14.7 | 8.4 |
| ≥ 0.8 | 20.1 | 13.4 |

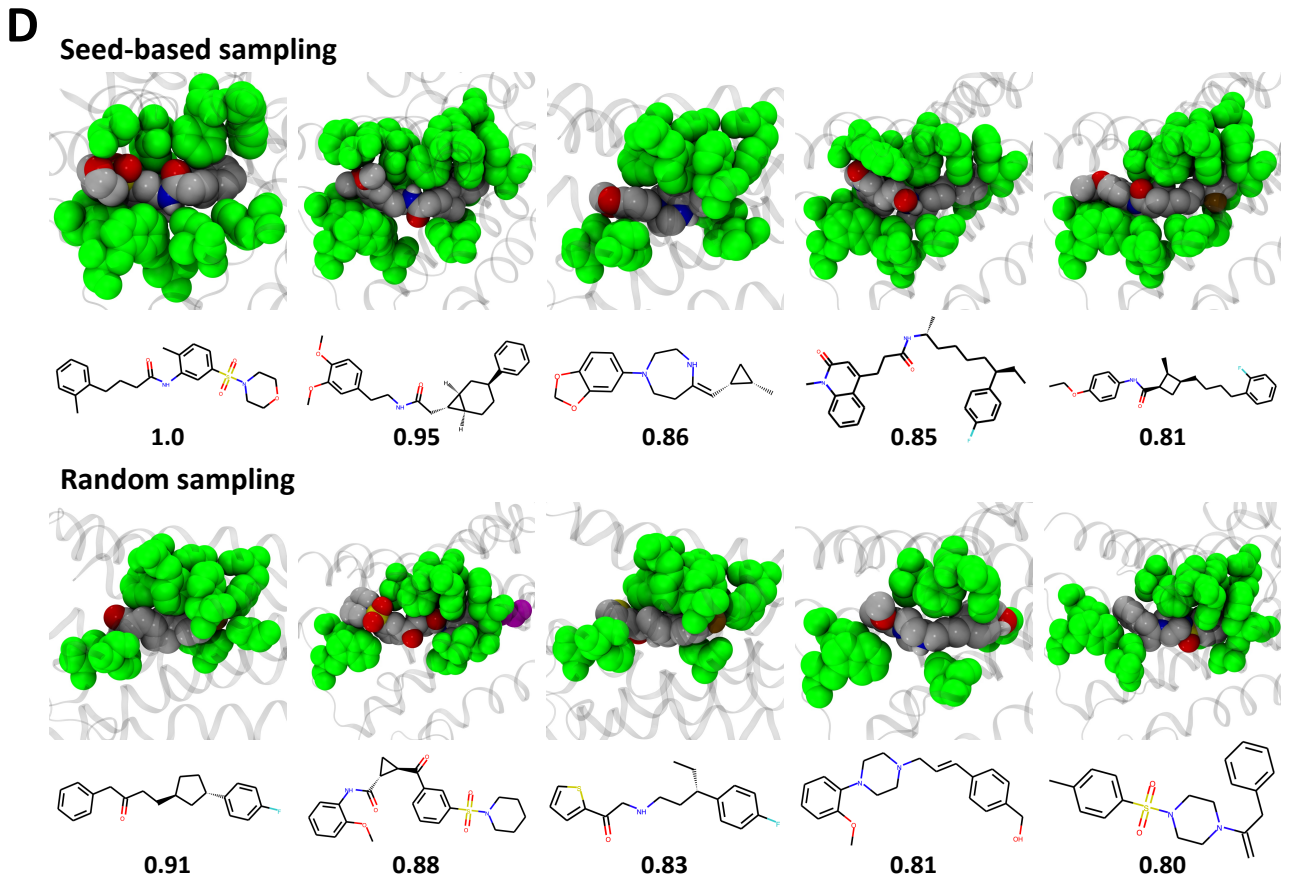

**Figure 2**. Docking, stereoisomer, and synthetic chemical database analysis. (A) Docking scores for gen3D and gen2D molecules from seed-based and random sampling, including average docking scores for the top 3 poses (left) and average differences between active and inactive conditioned molecules (right). (B) Distribution of stereoisomeric center count for molecules generated from gen3D and gen2D protocols. (C) Recovery rate of randomly-sampled generated molecules within the Enamine real database. (D) Docking-generated binding poses of selected molecules produced by the gen3D model. Generated molecules are shown in classic element colors, and generative-predicted contacting residues are shown in green.

Our gen3D model also produced more desirable results in terms of stereochemistry. The gen3D procedure yields stereocenter counts more consistent with reference seeds and the training set at large across both generation protocols and activity conditions; by contrast, the gen2D approach tends to favor higher stereocenter counts that complicate both synthesis and conformational selection (Fig 2B). It is intriguing that our gen3D method can implicitly learn to match stereocenter counts in reference data and suggests that the constraints imposed by training on protein-ligand complexes restrict configurational diversity of generated molecules to shapes complementary to the target's binding pocket, thereby improving docking scores (Fig 2. A and D).

Despite significant differences in seed-based docking results between gen3D and gen2D models, most physicochemical properties (e.g., molecular weight (MW), quantitative estimate of drug-likeness (QED), and synthetic accessibility (SA)) of generated molecules were similar in both procedures. Interestingly, the gen3D model tends to generate molecules that are relatively easy to synthesize - having lower SA - possibly owing to stereochemical complexity[32,33] (see SI for more discussion). Notably, molecules generated via gen3D were somewhat more hydrophobic than their gen2D counterparts, matching the generally hydrophobic nature of the DD2R ligand binding pocket (see SI for more discussion).

Consistent with these findings, the gen3D approach produces more molecules (same or similar) that appear in an Enamine library[34] search: 1.5 to 1.8 times more randomly sampled gen3D molecules were recovered in an Enamine search than gen2D molecules. Training on 3D protein-ligand complexes thus appears to yield more chemically accessible molecules, possibly owing to the realistic structural constraints imposed by the protein target binding pocket[32,33].

The gen3D model yields contact pairs, not coordinates. To what extent do protein-ligand contacts predicted by our generative model match those observed in docking poses? To address this question, we calculated the contact recovery rate (contact cutoff = 4.5 Å) for each generated molecule, as defined by $|C_{Dock}(m) \cap C_{Gen}(m)|/|C_{Gen}(m)|$, where the $C_{Gen}(m)$ and $C_{Dock}(m)$ are the contact-pair (i.e., ligand atom and contacting residue) sets for a molecule m from the generative prediction and the docking structure, respectively; the vertical lines $|\cdot|$ represent a cardinality operation. Predicted contacts and poses for an exemplar molecule are shown in Fig. 3A.

Our recovery rate results are divided into categories associated with the "first" and "best" docking poses for each generated molecule. The "first" pose is the binding mode with the best docking score among all docking poses generated for all enantiomers of the molecule. Selecting poses in this fashion, the contact recovery rate for first poses is distributed bimodally in both sampling protocols (Fig 3B). A majority or near-majority of molecules are found in regions of high contact recovery rate: first poses for ~60% of seed-based- and ~50% of randomly sampled- molecules recovered more than 50% of docking contacts.

The docking score is an imperfect measure of pose correctness and lower-ranked docking poses may also be physically correct.[35,36] Our "best" pose criterion thus first identifies the pose with the single highest recovery rate among the top ten poses for each molecular enantiomer, and subsequently selects the highest-scored enantiomer among these best-recovered poses. The low-recovery peak disappears in the "best" pose distributions for both generation protocols (Fig 3B). Indeed, best poses for more than 90% of seed-based molecules and 85% of randomly-sampled molecules were found to recover at least 50% of the predicted contacts. Thus, nearly all binding modes predicted by our 3D generative model were sampled by our docking program within a reasonable recovery rate.

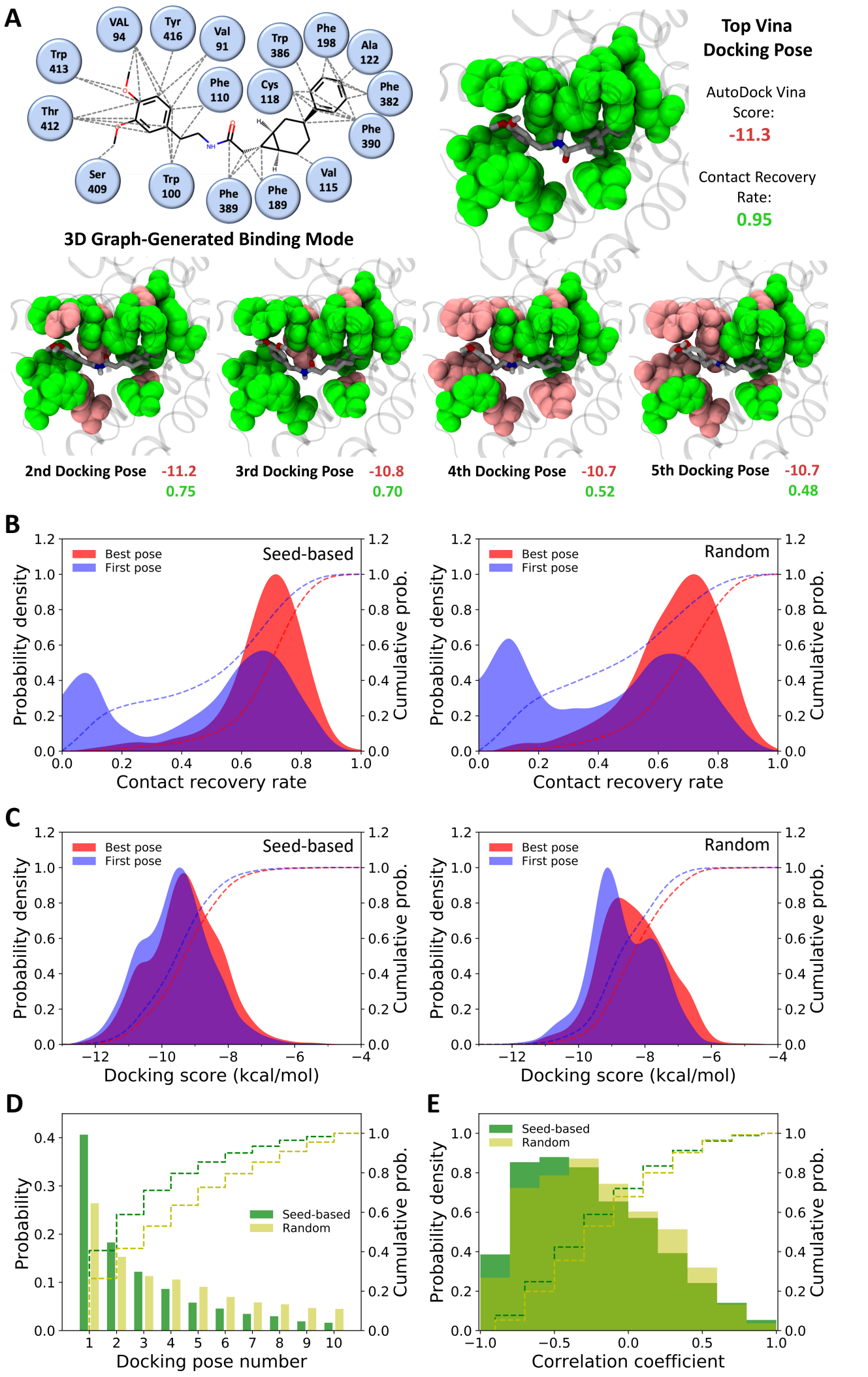

**Figure 3**. Recovery rates of docking pose contacts. (A) Docking poses and contact recovery rates for an example gen3D molecule (green residues have one or more recovered contact). (B) Contact recovery distribution between docking and gen3D-predicted poses, for both the highest recovery ("best") pose and the top docking score ("first") pose. (C) Distribution of docking scores for best and first poses. (D) Population of best poses among the top 10 docking poses, as determined by docking score. (E) Distribution of Spearman rank-order correlation coefficients between docking scores and recovery rates for each molecule's top 10 docking poses.

The docking score distribution of the first pose approximates that of the best pose, especially in the seed-based case (Fig 3C), suggesting that poses with the best recovery rate are typically not far from the top scoring pose. More specifically, the relative populations of selected poses as a function of docking pose rank indicate that the top pose most frequently exhibits the highest contact recovery in both generation protocols. Populations of highly recovered poses gradually decline toward lower docking pose rank. More than 85% of seed-based molecules featured a "best" pose within the top five docking poses; the same observation holds (though is slightly diminished, at 73%) among randomly sampled molecules (Fig 3D).

These data indicate a correlation between the docking score and the contact recovery rate. To explicitly measure this effect, we computed Spearman rank-order correlation coefficients between docking score and corresponding contact recovery rates for the top ten poses of each stereoisomer. As shown in Fig 3E, a majority of molecules (~72% seed-based/68% random) were found be anticorrelated on these metrics, meaning that the more negative (i.e., better) the docking score, the higher the contact recovery rate. This plot quantifies the notion that our gen3D model can successfully learn a generated ligand's suitability for a specific binding pocket, as can be captured with docking simulations. These contact recovery results suggest that our 3D generative approach can serve as a rough binding mode prediction tool and surrogate for docking simulations[37]; additional applications will be explored in future works.

Our results demonstrate that training generative models explicitly on 3D protein-ligand complexes enhances small molecule compatibility with the intended target's binding pocket. The graph-based gen3D approach predicts putative binding modes of generated molecules in a manner consistent with docking results, capturing the structural context of the ligand binding site. These 3D generative methods have thus begun to learn the underlying physics that govern protein-ligand binding, an important early step in accelerating structure-based molecular design for drug discovery. Specifically, this work highlights the broad applicability of graph-based deep learning architectures for modeling interacting chemical systems. While docking simulations served as our primary mode of validation in this work, more rigorous tests of our approach (both in-silico and in-vitro) and predicted drug-target interactions are welcomed by the authors.

## ASSOCIATED CONTENT

### Supporting Information

The Supporting Information describes the theory underlying our graph-based 3D generative models, procedures for model hyperparameterization and training, the method of standardization for our training dataset, molecular generation protocols, analysis of molecular properties, docking simulation protocols, the impact of JSD weights on docking and contact recovery results, and the chemical environment of DD2R binding site.

## AUTHOR INFORMATION

### Corresponding Author

* Email: sgkang@us.ibm.com

### Author Contributions

†These authors contributed equally.

## ACKNOWLEDGMENT

We thank Sugato Bagchi for assistance in training dataset preparation.

# Supplemental Information

This Supplemental Information describes the theory and implementation of the generative model presented in the main text, particularly with respect to how that model combines 3D protein-ligand contact networks with architectures and concepts associated with conditional-variational autoencoders (c-VAEs). Specific topics include the model's graph-based encoder and decoder, latent space, loss function, and training and optimization protocol. We also report the hyperparameterization for the 3D model, molecular properties of the 2D and 3D generative models, the docking and the contact recovery rates for the 3D models, and characteristics of the DD2R binding site.

## Construction of relational-graphs for small-molecules and protein-ligand contacts

Mathematical graphs provide a natural molecular representation, with nodes corresponding to atoms and edges to bonds. In previous work[1], we represented isolated ligands as molecular graphs; here, we expand this representation to include protein-ligand contacts. Given an arbitrary protein-ligand complex, ligand atoms ($N_{atoms} \leq N_{max}$, where $N_{max}$=40) were represented with one-hot encoded node vectors of size $N_{atoms} \times N_A$, with the second dimension associated with element type ($N_A$= 10, where $A = \{a|\ a \in (C,\ N,\ O,\ F,\ S,\ Cl,\ Br,\ I,\ \emptyset)\}$). The target activity ($f$) of ligands was concatenated as a binary activity condition to the atomic representation vector. Bonds between atoms were represented by stacking adjacency matrices, each constructed according to bond type ($N_B$= 5, where $B = \{b|\ b \in (single,\ double,\ triple,\ aromatic,\ \emptyset)\}$) to yield a relational graph of $N_B \times N_{atoms} \times N_{atoms}$.

Protein-ligand contacts were also described *via* a graph representation, with nodes corresponding to either ligand atoms or protein residues and edges describing their interactions. Residue nodes were included for amino acids within the DD2R binding site ($N_{residues} = 26$) and combined with one-hot encodings of amino acid types ($N_{AA} = 20$), resulting in a tensor of size $N_{residues} \times N_{AA}$. Protein-ligand contacts were defined between ligand atoms and binding-site residues using a 4.5 Å distance cutoff; contact tensors assumed dimensions of of $N_C \times N_{atoms} \times N_{residues}$, with the first dimension representing interaction type ($N_C$= 7). Options for interaction type included: $C = \{c|\ c \in (aliphatic,\ aromatic,\ charged_{+/-},\ charged_{-/+},\ hbond_{A/D},\ hbond_{D/A})\}$, where $charged_{+/-}$ ($charged_{-/+}$) corresponds to a charged interaction between a positive (negative) ligand atom and negative (positive) residue atom, and $hbond_{A/D}$ ($hbond_{D/A}$) represents hydrogen bonding between a ligand acceptor (donor) and residue donor (acceptor). If one ligand atom made multiple interactions with the same protein residue, only the shortest-distance interaction among them was included; each ligand atom was allowed to have multiple interactions with different residues without further restraint.

## Graph-encoder for small-molecules and protein-ligand contacts *via* relational g-CNN

Input molecular features were encoded through two relational graph-convolutional neural networks[2]: one for ligands (the ligand-only component) and the other for protein-ligand contacts. For the ligand-only component, the input molecular graph was convolved as a function of each bond type:

$$h_i^{LL(l+1)} = \sigma\left(\sum_{b\in B}\sum_{j\in n_i^{LL}(b)}\frac{1}{N_i^b}W_b^{LL(l)}h_j^{LL(l)} + W_s^{LL(l)}h_i^{LL(l)}\right) \qquad \text{Eq.1}$$

where the $i^{\text{th}}$ hidden node of the $l^{\text{th}}$ hidden layer ($h_i^{LL(l)}$) is convolved with its neighbor nodes ($h_j^{LL(l)}$) after a dense transformation ($W_b^{LL(l)}$) over all available neighbor nodes (indexed by $n_i^{LL}(b)$), connected through a bond type, $b$, and normalized by the number of connected neighbors ($N_i^b$). The convolved neurons were then activated with *tanh* functions, $\sigma(\cdot)$, after adding a self-connection term ($W_s^{LL(l)} h_i^{LL(l)}$). In the current encoder, graph convolutions were carried out to a depth of two, with each convolutional operation preceded by a respective dense layer (128 and 64 dimensions). Following another 128-dimensional dense transformation, nodes were aggregated by summing over an element-wise product of sigmoid and tanh activations along the bond type in order to produce an embedding invariant to atom order. Two subsequent layers corresponding to dense transformations (128 and 64 dimensions) were then used to generate the final output for the ligand-only component.

Meanwhile, the protein-ligand contact graph was encoded via convolutional operations among ligand nodes and protein residue nodes within the relational protein-ligand contact tensor as a function of interaction type (C), as follows:

$$h_p^{CL(l+1)} = \sigma\left(\sum_{c\in C}\sum_{q\in n_p^{CP}(c)} W_c^{CP(l)} h_q^{CP(l)} + W_s^{CL(l)} h_p^{CL(l)}\right) \qquad \text{Eq.2}$$

$$h_u^{CP(l+1)} = \sigma\left(\sum_{c\in C}\sum_{v\in n_u^{CL}(c)} W_c^{CL(l)} h_v^{CL(l)} + W_s^{CP(l)} h_u^{CP(l)}\right) \qquad \text{Eq.3}$$

Eq.2 represents the relational graph convolutional neural network for ligand nodes across contacting protein residues, as defined in the protein-ligand contact tensor; Eq.3 represents the complementary neural network for protein binding-site residues across contacting ligand atoms.

On the ligand side (Eq.2), $h_p^{CL(l)}$, the $p^{th}$ hidden node of the $l^{th}$ layer for a ligand atom $p$ was convolved with protein nodes ($h_q^{CP(l)}$), after a dense transformation ($W_c^{CP(l)}$) over all available contacting residue nodes ($n_p^{CP}(c)$) of an interaction type $c$; these nodes were then subjected to *tanh* activations ($\sigma(\cdot)$) after addition of the self-connection term for the ligand atom ($W_s^{CL(l)} h_p^{CL(l)}$). On the residue side (Eq.3), $h_u^{CP(l)}$, the $u^{th}$ hidden node of the $l^{th}$ layer for a residue $u$ was convolved with ligand nodes ($h_v^{CL(l)}$), after a dense transformation ($W_c^{CL(l)}$) over all available contacting ligand atoms ($n_u^{CL}(c)$) of a interaction type $c$; these nodes were similarly activated with a *tanh* function after addition of the self-connection term for the residue ($W_s^{CP(l)} h_u^{CP(l)}$).

This procedure allowed nodes from both ligand atoms and protein residues to be intertwined through the relational tensor of protein-ligand contacts. As in the ligand-only component, convolutions were performed to a depth of two, each with respective dense operations with dimensions of 128 and 64) preceding the convolution operation (i.e., $W_c^{CP(l)}$, $W_s^{CL(l)}$, $W_c^{CL(l)}$ and $W_s^{CP(l)}$).

Once generated, the ligand and protein residue sides of the protein-ligand contact representation were aggregated by summing over all contact types after an element-wise multiplication of sigmoid and tanh activations succeeding another 128-dimension dense transformation on the hidden nodes. The final output for the protein-contact representation was produced by concatenating the aggregated tensors from both sides, followed by two additional dense layer operations (128 and 64 dimensions).

### Molecular latent space with protein-contact and activity conditioning

The graph encodings were then projected onto variational parameters of means and standard deviations of given dimensions to facilitate latent space sampling. The ligand-only encodings were mapped over means ($\boldsymbol{\mu}_{LL}$) and standard deviations ($\boldsymbol{\sigma}_{LL}$), with the embedding taking the shape of a 32-dimensional vector that matches our previous model developed for isolated ligands[1]. Meanwhile, encodings of protein-ligand contacts were mapped onto separate variational parameters ($\boldsymbol{\mu}_{PL}$ and $\boldsymbol{\sigma}_{PL}$), where the embedding dimension was selected from among values of 8, 16, and 32 as a hyperparameter to be optimized. Next, the component embeddings were concatenated to form joint embeddings $\boldsymbol{\mu} = \boldsymbol{\mu}_{LL}||\boldsymbol{\mu}_{PL}$ and $\boldsymbol{\sigma} = \boldsymbol{\sigma}_{LL}||\boldsymbol{\sigma}_{PL}$, from which the latent space vectors, $\boldsymbol{z}' = \boldsymbol{\mu} + \boldsymbol{\epsilon} \odot \boldsymbol{\sigma}$ were sampled using a Gaussian noise parameter $\boldsymbol{\epsilon} \sim \mathcal{N}(\mathbf{0}, \mathbf{1})$.

Lastly, the binary target activity conditions (*f*) were attached to $\boldsymbol{z}'$ in order to generate the final latent embedding vector $\boldsymbol{z}$, thus resulting in a latent representation of 42, 50, or 66 dimensions depending on the choice of embedding dimension for the protein-ligand contact component. Latent space sampling was performed by obtaining embeddings for points either (1) surrounding encoded molecular seeds or (2) directly sampled at random within the latent space using no seeds.

### Graph-decoder for generating molecular graphs and predicting protein-ligand contacts

Our graph decoder was designed to generate relational molecular graphs for both small-molecules and their cognate putative binding modes. Thus, a ligand would be decoded into node vectors and an edge tensor for atoms and bonds, respectively, and a ligand-binding mode into an edge tensor for possible interactions between the ligand and protein binding-site residues. Decoding began with the transformation of the latent vector $\boldsymbol{z}$ with two fully connected dense layers of 64 and 128 dimensions and *tanh* activation functions, branching the latent representation into two sets, each expanded to the desired sizes for atom ($N_{max}$ = 40) and residue ($N_{residues}$ = 26) nodes, respectively. The first branch was utilized for decoding small molecular graphs, and the second branch decoded the protein-ligand binding network. Each branch was then independently plugged into a dynamic recurrent neural network (RNN) containing 300 LSTM cells to produce RNN outputs.

The RNN output from the ligand component was then linearly transformed to create output logits for nodes ($N_{max} \times N_A$) and edges ($N_{max} \times N_{max} \times N_B$) with no activation function, while atomic features (aliphatic and aromatic rings) were simultaneously decoded with 8- and 4-dimensional dense transformations with tanh activation ; a final 2-dimensional linear transformation was then applied with no activation. Meanwhile, the protein-ligand contact component was decoded using linear transformations for both ligand-side and protein-side RNN outputs; those outputs were then mixed using a simple matrix multiplication to produce a logit tensor representing protein-ligand contacts ($N_{max} \times N_{residues} \times N_C$).

## Enhancement of protein-ligand interaction integrity

Trained with 3D protein-ligand complex data, our generative model was designed to generate not only new molecules, but also to predict putative binding modes between generated ligands and their respective target. For the latter task, we introduced an additional function to enhance the fidelity of predicted interactions between the ligand and binding-site residues. Specifically, a Jensen-Shannon divergence (JSD) was employed to capture similarities between probability distributions for ligand and protein residues over all available interaction types:

$$D_{JSD} = \frac{1}{N}\sum_{i=1}^{N} JSD(p_i, q_i) \qquad \text{Eq.4}$$

, where

$$p_i(c) = \frac{1}{\mathcal{N}_i^{LL}}\sum_{a\in c} p_i^{LL}(a) \qquad \text{Eq.5}$$

$$q_i(c) = \frac{1}{\mathcal{N}_i^{PL}}\sum_{j\in R} q_{i,j}^{PL}(c) \qquad \text{Eq.6}$$

Here, the $JSD(p_i, q_i)$ measures a distance between two probability distributions: the first ($p_i$) constructed from the node prediction for ligand atoms and the second ($q_i$) from the edge prediction for protein-ligand contacts. The distribution $p_i$ of a ligand atom $i$ indicates the probability ($p_i(c)$) of an atom $i$ engaging a protein residue via a particular type of interaction $c \in \{aliphatic,\ aromatic,\ charged_{+/-},\ charged_{-/+},\ hbond_{A/D},\ hbond_{D/A}\}$ from a node prediction standpoint.

Thus, the distribution $p_i(c)$ is composed of node prediction probabilities $p_i^{LL}(a)$ that are summed over atom types ($a$), grouped into the possible interaction types $c$, and normalized by $\mathcal{N}_i^{LL} = \sum_c p_i(c)$. The distribution $q_i$ of a ligand atom $i$ represents the probability ($q_i(c)$) of an atom $i$ interacting with protein binding-site residues via the interaction type $c$ from the standpoint of protein-ligand contact edge prediction. The $q_i(c)$ can therefore be prepared by summing the contact probabilities $q_{i,j}^{PL}(c)$ for the interaction type $c$ between the ligand atom $i$ and a protein residue $j$ over all available residues in $R$ (the protein binding pocket) and normalizing with $\mathcal{N}_i^{PL} = \sum_c q_i(c)$. The overall distance $D_{JSD}$ for an $N$-atom molecule was then obtained by averaging over all atomic *JSD* values. This JSD distance was added to the loss function and scaled by a hyperparameter to be optimized.

## Loss function for the model optimization

Models were trained to generate the posterior distribution $q_\phi(z|x, f)$ for the latent space variable $z$ and probability distribution $p_\theta(x|z, f)$ for the molecular graph $x$ using at a given target activity condition $f$ applied to both the encoder ($\phi$) and decoder ($\theta$) networks. Thus, the evidence lower bound (ELBO) for our models was defined as follows:

$$ELBO = \mathbb{E}_{q_\phi(z|x,f)}(\log p_\theta(x|z, f)) - KL(q_\phi(z|x, f)\|p(z)) \qquad \text{Eq.7}$$

The first term in this equation represents the sum of log-likelihoods for reconstruction of the ligand nodes, ligand edges, protein-ligand contact edges, and ligand atomic features. The second term is the regularizing Kullback-Leibler (KL) divergence between the Gaussian prior, $p(z)\sim\mathcal{N}(\mathbf{0}, \mathbf{1})$, and the posterior, $q_\phi(z|x,c)$ drawn from the encoder. Adding the $JSD$ term, we obtain the loss function $L$ to be minimized:

$$L = -\mathbb{E}_{q_\phi(z|x,f)}(\log p_\theta(x|z,f)) + \beta KL(q_\phi(z|x,f)\|p(z)) + \gamma \overline{D}_{JSD} \qquad \text{Eq.8}$$

As noted above, the batched averaged $JSD$ term for protein-ligand contact integrity, $\overline{D}_{JSD}$, is scaled by an adjustable weight factor $\gamma$. The KL contribution was also combined with an adjustable weight factor $\beta$, as seen in $\beta$-VAEs[3]. A smaller $\beta$ in early epochs allows the network to be more efficiently optimized in parameter space with respect to the molecular graphs, while gradually growing more regularized at later stages of training.

**Training database for 2D and 3D generative models**

High-throughput screening assay data from PubChem Assay ID 485358[4] were employed for both 2D (gen2D) and 3D (gen3D) generative models. The dataset contains 1,780 active-labeled and 324,800 inactive-labeled molecules screened against the human dopamine D2 receptor (DD2R), from which the training datasets were prepared after our standardization protocol.

Molecules were excluded if they met one of the following conditions: (i) molecules composed of >40 atoms, (ii) compounds adducted with infrequently occurring ions/atoms (e.g., $H^+$, $Li^+$, $B^-$, $Na^+$, Si, K+, $Ca^{2+}$, $Zn^{2+}$, As, Se, $Ag^+$, $Nd^{3+}$, $Pt^{2+}$, Hg), (iii) organometallic compounds (e.g., Fe-coordinated ferrocene derivatives), and (iv) compounds adducted with organic/inorganic acids (e.g., potassium hydrogenoxalte, pyridine, maleic acid, fumaric acid, orotic acid, benzoic acid, succinic acid, citric acid, derivatives of pyridinecarboxylic acid, picric acid, hydrochloric acid, bromic acid, iodic acid, nitric acid, phosphoric acid, sulfuric acid, percloric acid, methanesulfonic acid, etc.). These exclusion criteria resulted in remainders of 1,505 active and 323,711 inactive molecules, respectively. We prepared two different training sets from the remaining data: (1) a smaller set for gen3D hyperparametization composed of all active molecules and 25,000 randomly selected inactive molecules (reduced dataset), and (2) a larger set for comprehensive gen3D and gen2D modeling composed of all active molecules and a randomly selected half (i.e., ~150,000) of the inactive molecules (full dataset).

The selected molecules were then canonized into a SMILES representation followed by conversion to MOL file format using RDKit (an open-source toolkit for cheminfomatics)[5]. A graph featurization protocol generated node and edge tensors for each molecule using RDKit definitions for atom-types (i.e., C, N, O, F, P, S, Cl, Br, or I), bond-types (i.e., Chem.rdchem.BondType.[ZERO, SINGLE, DOUBLE, TRIPLE, or AROMATIC]), and additional atomic features (whether an atom is in a ring and whether that ring is aromatic). All molecular featurization was done with standard functions from RDKit combined with the open-source python libraries pandas and NumPy.

For the gen3D training sets, we obtained 3D structures for the selected ligands bound in the binding pocket of the DD2R using the docking program AutoDock Vina[6]. The starting 3D structures of the ligands were obtained from their SMILES representations using the RDKit library. Molecules were then docked into the protein structure of DD2R obtained from PDB entry 6CM4 using a cubic search box (L=27 Å) centered on the binding pocket of the ligand labeled 8NU in that entry. Up to 10 binding modes were obtained from docking.

## Hyperparameterization for the 3D generative model

The gen3D architecture was based in part on our previous gen2D model[1], which was here augmented with a 3D protein-ligand contact graph and JSD constraints. Preserving hyperparameters optimized for the gen2D model, two additional hyperparameters within the gen3D remained to be optimized: (1) the additional latent space dimension for the protein-ligand contacts and (2) the weight factor $\gamma$ for the JSD term ($D_{JSD}$) that maintains protein-ligand contact integrity. We conducted a grid search across protein-ligand latent space dimensions of 8, 16, and 32 and JSD weight factor $\gamma$ values of 0.0, 0.5, 1.0, and 2.0, yielding the 12 different combinations shown Fig. S1.

*JSD* weight factor, $\boldsymbol{\gamma}$

| Embedding dim. $d_{PL}$ | 0.0 | 0.5 | 1.0 | 2.0 |
|---|---|---|---|---|
| 8 | **hp1f** | **hp2f** | **hp3f** | **hp4f** |
| 16 | **hp5f** | **hp6f** | **hp7f** | **hp8f** |
| 32 | **hp9f** | **hp10f** | **hp11f** | **hp12f** |

**Figure S1**. Hyperparameter grid-search for gen3D architecture. The embedding dimension for the protein-ligand contacts was set to 8, 16, or 32, while the JSD ranged from 0.0 to 2.0.

Four different gen3D models were independently trained for each hyperparameter condition for 1,000 epochs over the reduced training dataset. Model performance was assessed using generated molecule yield, novelty, and uniqueness parameters averaged over the four models (Fig S2). The yield measured how many valid molecules (after RDKit sanitization) were generated from a given number of trials; the novelty measured how novel the generated molecules were compared to the training dataset; and the uniqueness measured how unique each generated molecule was compared to other generated molecules. Such considerations yielded two preferred hyperparameter sets for more rigorous modeling with the full training dataset: $d_{PL}$ = 8 and $\gamma$ = 1.0, and $d_{PL}$ = 16 and $\gamma$ = 2.0, named hp3f and hp8f, respectively. The latter (hp8f) was selected as the featured hyperparameter set for this work, referred to simply as gen3D in the main text.

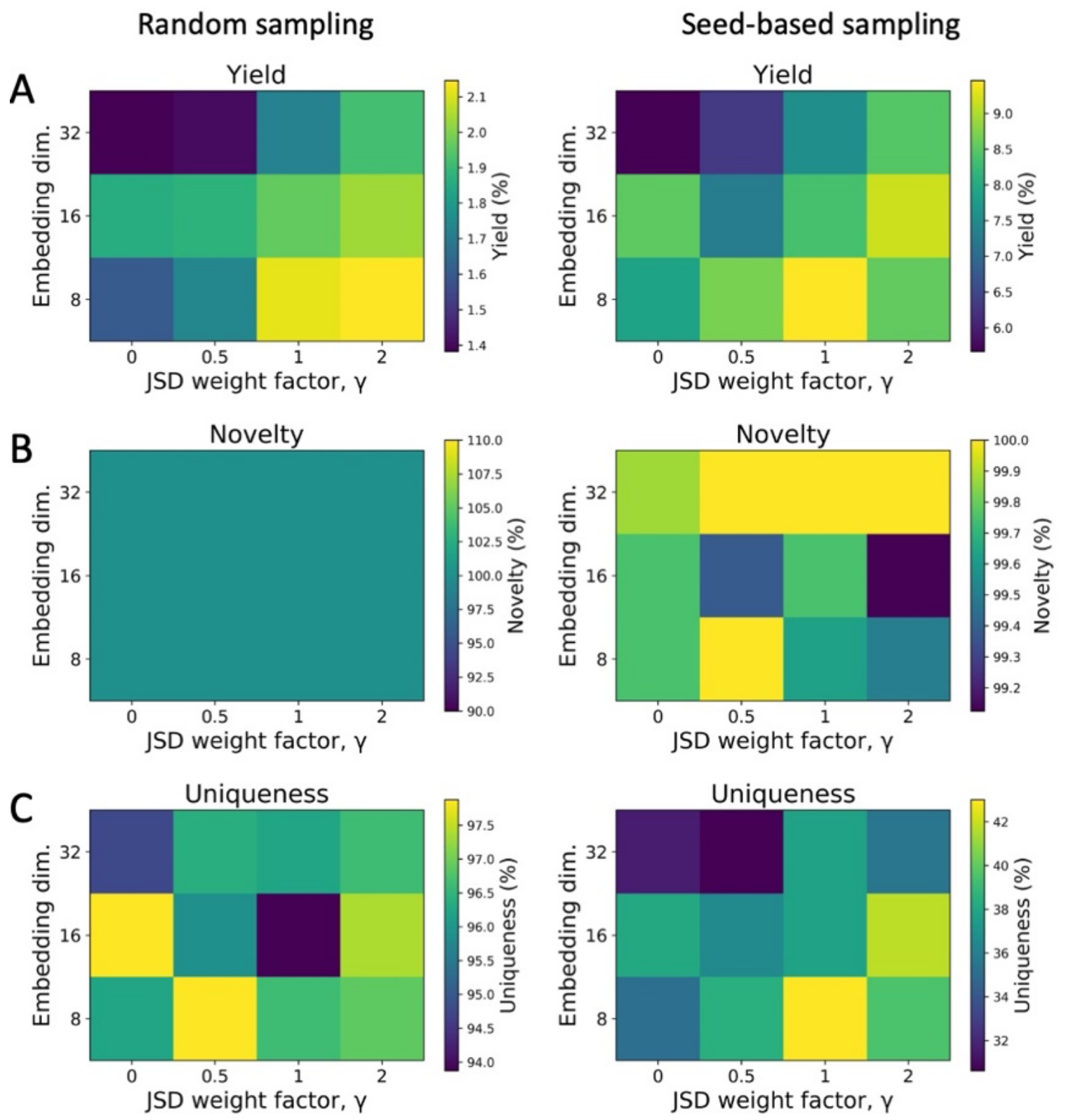

**Figure S2**. Model performance as a function of hyperparameters. In each generation protocol, yield (A), novelty (B) and uniqueness (C) metrics were applied to each combination of hyperparameters.

## Model Training

All generative models were trained under the framework of the TensorFlow Python library[7]. Model weight parameters were optimized with the Adam optimizer[8] with default learning rate = 0.001, $\beta_1$ = 0.9, and $\beta_2$ = 0.999 settings; weights were updated after every batch (each containing 200 data points) over 2,000 epochs for training with the full dataset for the selected hyperparameters (i.e., hp3f and hp8f). As mentioned above, we began the model training with no contribution from the KL term to the loss (Eq. 8), thus giving more initial weight to the reconstruction loss; the KL term's contribution was gradually increased by varying the weight factor $\beta$ linearly from 0 to 1 over the first 21 epochs. This same protocol was applied in training the 2D generative model. Loss curves for each model are shown in Fig. 3S.

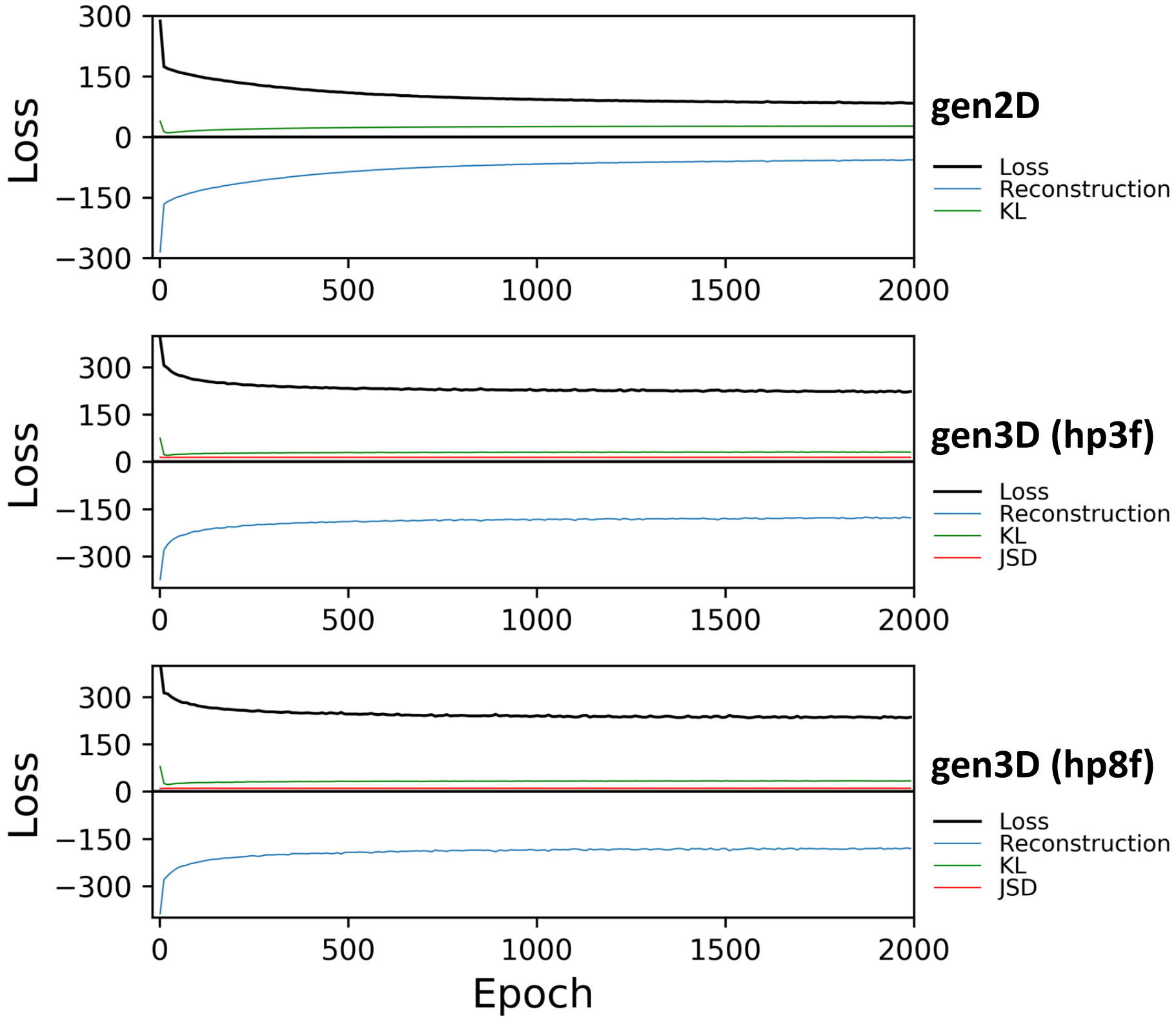

**Figure S3**. Loss changes over the course of 2D and 3D generative model training. In each model, the loss function (black curve) is shown with constituent terms of reconstruction (blue), KL divergence (green), and JSD (red).

## Generation protocols for molecules

As noted above, our models were also trained on molecular activity conditions. Two protocols were applied for generating molecules with a desired activity: (1) direct sampling from the latent space (random sampling) and (2) seed-based sampling specific to input molecules (seed-based sampling). For the former procedure, we constructed the latent space embedding $z$ by first directly sampling from Gaussian noise $z' \sim \mathcal{N}(\mathbf{0}, \sigma \cdot \mathbf{1})$ (where $\sigma$= 1.0 or 2.0) set to a latent space dimension (e.g., 48 for gen3D (hp8f) and 32 for gen2D) and then concatenating an activity condition ((1,0) or (0,1) for the inactive and active conditions, respectively). In seed-based sampling, a seed molecule and corresponding activity were fed through the encoder to produce variational parameters (mean and standard deviation of the seed), around which a Gaussian noise parameter $\boldsymbol{\epsilon} \sim \mathcal{N}(\mathbf{0}, \sigma \cdot \mathbf{1})$ (where $\sigma$ = 1.0 or 2.0) was sampled and then concatenated with the activity condition to generate the embedding for the decoder.

Numbers of molecules and their associated stereoisomer totals sampled across various protocols are included in Table S1. For each combination of activity conditions, sampling standard deviation, and sampling protocols, a minimum of 1,000 generated molecules were produced. A threshold of 1,000 proved sufficient to differentiate mean docking scores at approximately 95% confidence of standard error in all relevant cases. Additional new and unique molecules were obtained when the process was expedient; up to nearly 20,000 molecules were sampled in some protocols.

**Table S1**. Number of generated molecules and corresponding stereoisomer totals.

| | | | 2D generative modeling | | 3D generative modeling | | | |
|---|---|---|---|---|---|---|---|---|
| | | | Gen2D | | hp3f | | gen3D(hp8f) | |
| | Cond.[a] | $\sigma$[b] | Number | Number of stereoisomers | Number | Number of stereoisomers | Number | Number of Stereoisomers |
| Random | 0 | 1.0 | 3,128 | 11,630 | 2,335 | 3,695 | 2,245 | 3,820 |
| | 0 | 2.0 | 2,111 | 17,438 | 2,171 | 4,712 | 1,910 | 4,510 |
| | 1 | 1.0 | 2,679 | 17,670 | 2,438 | 4,227 | 2,302 | 4,185 |
| | 1 | 2.0 | 1,792 | 16,961 | 2,253 | 5,204 | 1,966 | 5,876 |
| Seed-based | 0 | 1.0 | 5,403 | 31,652 | 3,720 | 8,340 | 4,361 | 11,371 |
| | 0 | 2.0 | 15,065 | 85,854 | 10,275 | 22,860 | 12,392 | 32,083 |
| | 1 | 1.0 | 7,482 | 39,005 | 5,446 | 15,071 | 5,881 | 35,687 |
| | 1 | 2.0 | 20,797 | 115,320 | 16,047 | 45,161 | 19,239 | 93,851 |

[a]The activity condition concatenated to the latent space vector.

[b]The sampling standard deviation for the Gaussian noise sampling

## Calculation for molecular properties

Molecular properties such as molecular weight (MW), quantitative estimate of drug-likeness (QED), synthetic accessibility (SA), and LogP of the training and generated data were computed with the RDKit cheminformatics library[5]. There, QED and SA scores were obtained based on Bickerton et al[9]. and Ertl et al[10].'s methods, respectively. LogP values were calculated based on Crippen's method[11]. Briefly, the QED metric assesses the drug-like properties of molecules, considering such quantities as molecular weight, octanol-water partition coefficient, number of hydrogen bond donors/acceptors, polar surface area, number of rotatable bonds, number of aromatic rings, and number of structural alerts of a set of orally absorbed and approved drugs. SA provides relative synthesizability estimates consistent with those scored by medicinal chemists by assessing fragmental contributions and applying a complexity penalty to a putative synthesis. Molecular similarities of the generated molecules to those in the Enamine library (downloaded from https://enamine.net/compound-collections/real-compounds/real-database)[12] were calculated using the Tanimoto similarity metric for Morgan circular fingerprints[13] of radius 2.

## Molecular properties of generated molecules

Figure S4 shows the properties of molecules produced by the 3D and the 2D generative models. For the 3D models, we present two results (i.e., hp3f and hp8f) trained with different hyperparameter sets. Previously, we reported that our gen2D models can learn molecular features that were not conditioned during training. For example, random generation produced molecules with distributions of QED, SA, and logP similar to those of the training data; while the average MW was reduced, possibly due to the difficulty of direct searches in the latent space, MW values were still consistent with those of drug-like molecules. Similar patterns were observed in the seed-based results, with the exception that MW distributions were maintained close to the seed.

With 3D protein-ligand contacts integrated into the architecture, the 3D generative models yielded similar patterns in molecular properties (e.g., for MW, logP, and QED). However, the synthetic accessibility (SA) score was lower, on average, for molecules produced by the gen3D model, indicating the 3D models generate molecules that are relatively easy to synthesize. This finding is interesting in the context of the increased recovery rate observed for the gen3D method (cf. Fig 2C) within the Enamine chemical library[12]. Inclusion of protein-ligand contacts during training perhaps imposes a target-tailored structural confinement, yielding molecules with fewer stereoisomeric centers (cf. Fig 2B) that are more synthesizable[10] and more likely to appear in databases like Enamine.

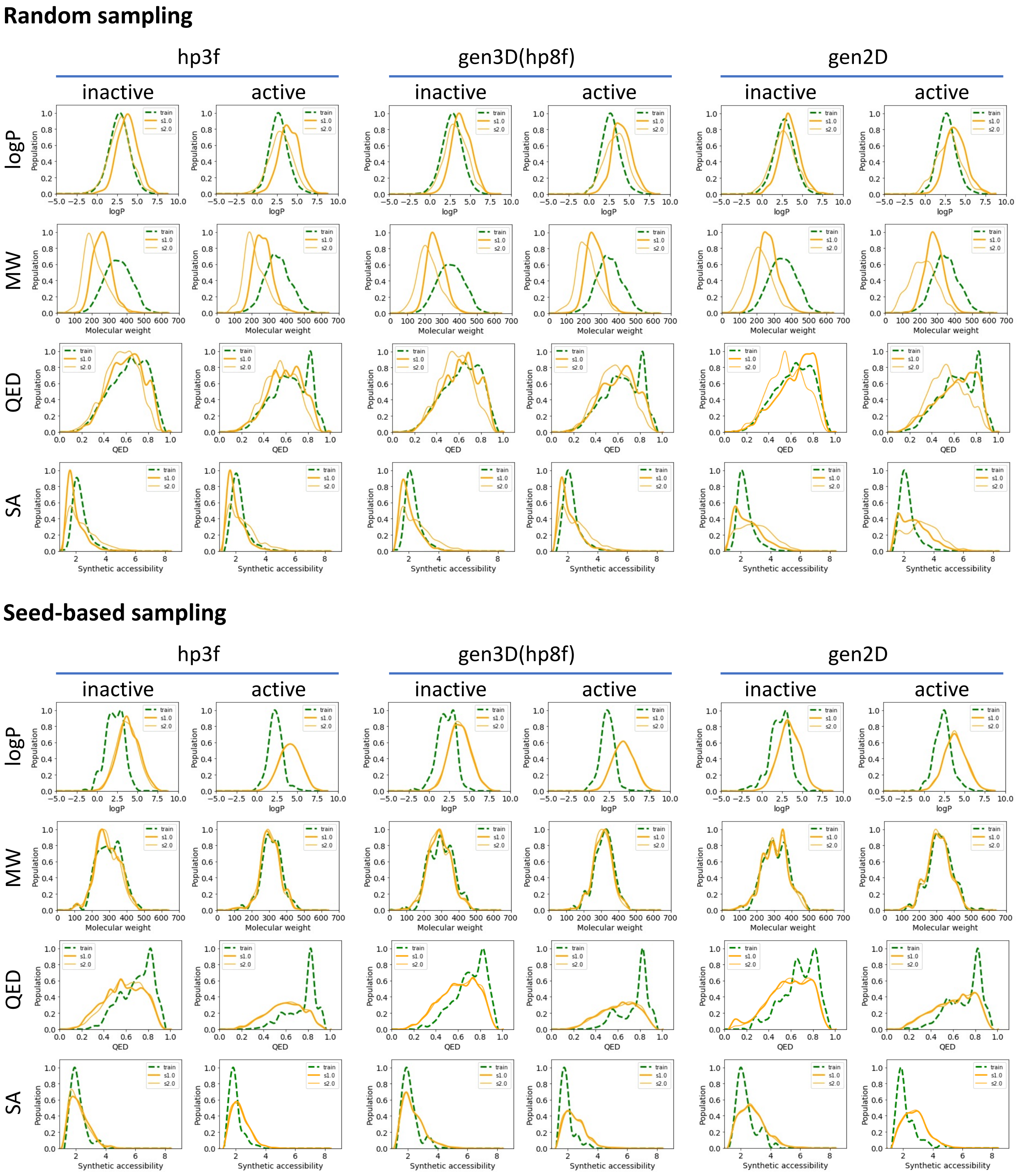

**Figure S4**. Physicochemical properties of generated molecules produced by 2D and 3D generative models. Results for two 3D generative models (hp3f and gen3D(hp8f)) and one 2D generative model (gen2D) are shown from the left to the right column for random (top) and the seed-based (bottom) sampling in each activity condition. The s1.0 and s2.0 labels indicate the sampling standard deviations of 1.0 and 2.0, respectively.

## Docking simulation for the generated molecules

The SMILES representation of each generated molecule was obtained from the ligand-only graphs output by the gen3D model by constructing a molecule object and then exporting to SMILES using RDKit functions. With SMILES representations in hand, the molecules were then prepared and docked into the protein structure of PDB entry 6CM4 using the same protocol outlined above for the docking of the training set ligands obtained from PubChem[4]. All available stereoisomers for molecules generated under each activity condition and sampling procedure were docked into the target binding site and assessed as described in the main text. A slightly higher exhaustiveness parameter (32 vs. 16) was applied in docking generated molecules.

## JSD effect on docking score and contact recovery

In order to enhance the integrity of the predicted protein-ligand contacts, we introduced the Jensen-Shannon divergence as an additional constraint (Eq.4) with a scaling factor $\gamma$, to be tuned by hyperparameterization. The hyperparameterization resulted in two reasonable options for $\gamma$: $\gamma$ = 1.0 (hp3f) or $\gamma$ = 2.0 (hp8f). Although we ultimately chose hp8f as our preferred hyperparameter set for 3D generative modeling (i.e., gen3D), we also analyzed the influence of the JSD contribution ($D_{JSD}$) on docking simulation results and protein-ligand binding mode predictions via comparisons of hp3f and hp8f.

As summarized in Fig S5A, increasing the JSD constraint (i.e., from hp3f to hp8f) was observed to enhance docking scores and recovery rates of the predicted ligand contacts. Specifically, the docking score distribution was shifted toward lower (better) scores in seed-based generation and weighted more to lower scores in the random sampling case. Similarly, the contact recovery rate trended toward higher values with greater JSD contribution (hp8f). Intriguingly, the contact recovery rate seems to be anticorrelated with the docking score, a trend which grows clearer when the docking score is normalized by the number of ligand atoms (Fig S5B). Although the correlation coefficients (i.e., $r$ = -0.3 ~ -0.2) are weak, $p$-values are quite low in all cases. These data indicate that although the correlation is not strong, a statistically significant correlation can be observed between lower docking scores and higher contact recovery rates.

Lastly, the distribution of the docking pose rank corresponding to the highest contact recovery mode indicates that higher-ranked docking poses tended to have higher contact recovery rates with a larger JSD contribution (hp8f), particularly in seed-based sampling (Fig S5C). This result demonstrates that a higher JSD term facilitates molecular generation that is more consistent with the binding-site structure, potentially enhancing the compatibility between the ligand and its target.

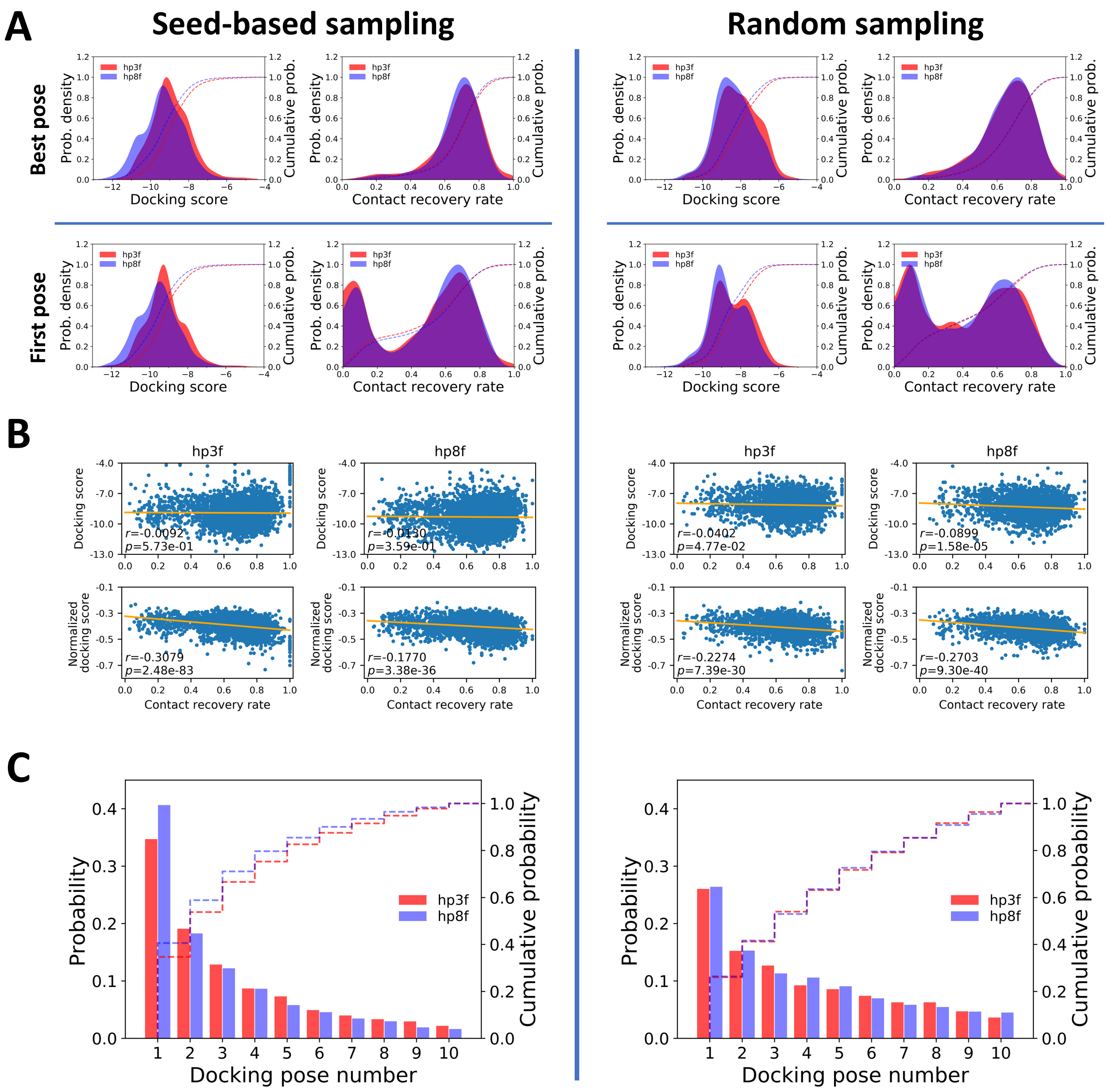

**Figure S5.** Effect of JSD weight on docking score and contact recovery rate. (A) Distributions of docking scores and contact recovery rates from seed-based (left) and the random (right) sampling. (B) Correlation between docking score and recovery rate of predicted protein residue contacts. (Top: docking scores vs. recovery rates, and bottom: normalized docking scores vs. recovery rates). (C) Distribution of the docking pose rank of the highest contact recovery pose as a function of JSD weight.

## Chemical environment of DD2R binding site

We also inspected the DD2R binding pocket to better understand the molecular properties that help determine ligand binding. Using the X-ray structure of DD2R co-crystalized with risperidone (PDB code: 6CM4) as a reference, we noted that the DD2R binding pocket is mostly composed of hydrophobic or aromatic residues (Fig S6); only a few polar residues and one charged residue (ASP114) were found. These pocket characteristics may imply the importance of hydrophobic and/or aromatic interactions for small molecule binding to DD2R at this location.

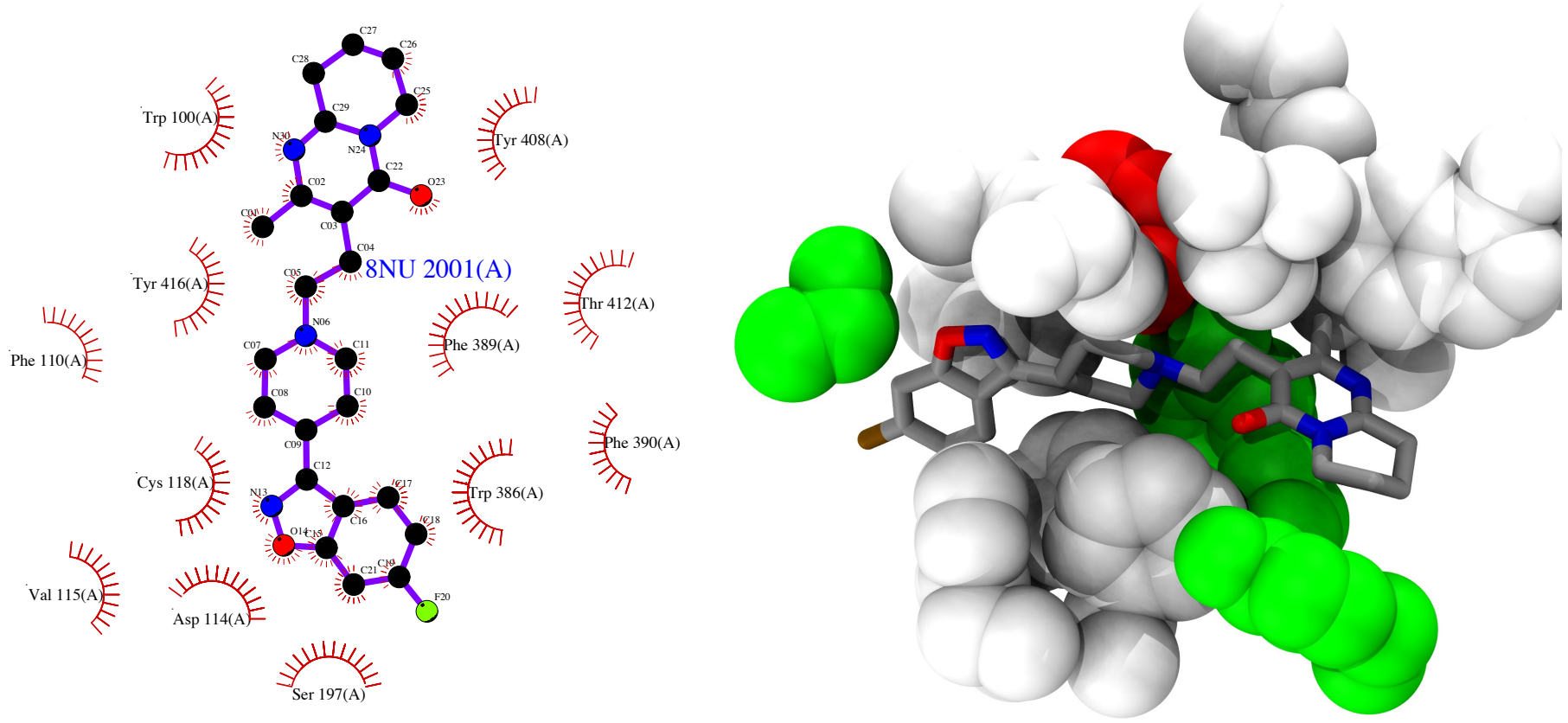

**Figure S6**. Binding site residues of the dopamine D2 receptor. 2D diagram of contact residues of DD2R on bound risperidone (left). Corresponding 3D structure of the DD2R binding site, where hydrophobic/aromatic, polar, and charged residues are colored in white, green, and red, respectively (right). The 2D diagram and 3D structure were drawn by LIGPLOT[14] and VMD[15], respectively.

## References for the Supplemental Information